\documentstyle[12pt,a4]{article}

\begin{document}

\setcounter{page}{0}

\begin{titlepage}

\hfill LAPTH-752/99

\vspace{2cm}

\begin{center}

{{\Large\bf
 Perturbation theory in radial quantization approach and the expectation
values of exponential fields in sine-Gordon model 
}}\\

\vspace{2cm}
{\large V.V.Mkhitaryan$^\dag$, R.H.Poghossian$^\ddag$, T.A.Sedrakyan$^\dag$} \\
\vspace{1cm}

{$^\dag$\ \em Yerevan Physics Institute,  }\\
{Alikhanian Brothers St. 2,  Yerevan 375036,  Armenia}\\
\vspace{5mm}

{$^\ddag$\ Laboratoire d'Annecy-leVieux de Physique Theorique LAPTH}\\
{LAPP, BP 110, F-74941 Annecy-le-Vieux Cedex, France;}\\
{permanent address: \em Yerevan Physics Institute,}\\
{Alikhanian Brothers St. 2,  Yerevan 375036, Armenia}
\end{center}

\vspace{2cm}
\centerline{{\bf{Abstract}}}

A perturbation theory for Massive Thirring Model (MTM) in radial 
quantization approach is developed. Investigation of the twisted sector in 
this theory allows us to calculate the vacuum expectation values of 
exponential fields $\left\langle \exp ia\varphi \left(0\right) \right\rangle$ 
of the sine-Gordon theory in first order over  Massive Thirring Models 
coupling constant. It appears that the apparent difficulty in 
radial quantization of massive theories, namely the explicite ''time'' 
dependence of the Hamiltonian, may be successfully overcome. The result we 
have obtained agrees with the exact formula conjectured by Lukyanov and 
Zamolodchikov and coincides with the analogous calculations recently 
carried out in dual angular quantization approach by one of the authors.    

\end{titlepage}

\vspace{0.5cm} \setcounter{equation}{0}

\section{Introduction}

\renewcommand{\theequation}{1.\arabic{equation}}

It is almost two decades the two dimensional exactly solvable 
models of QFT attract much attention because of their wide applications 
in the condenced matter physics and string theory. From the other hand, the 
knowlidge of the exact solutions of such nontrivial interacting theories 
provide us with better understanding of the general concepts of QFT.
While the complete on-shell solution (i.e. the mass spectrum and exact 
s-matrix) for many of 2d Integrable QFT'es (IQFT) are well known, 
the construction of form-factors and correlation functions are under 
current intensive ivestigations now.

The main subject of investigation in this paper is the most studied example of 
2d IQFT the sine-Gordon model, which is defined by the action
\begin{equation}
{\cal S}_{SG}=\int d^{2}x\left\{ \frac{1}{16\pi }\partial _{\nu }\varphi
\partial ^{\nu }\varphi +2\mu \cos \beta \varphi \right\}  \label{0.1}
\end{equation}                 
where $\varphi $  is a real Bose field. The spectrum of this model includes 
the soliton, anti soliton and some number (depended on the coupling constant
$\beta$) of their bound states named breathers. The scattering in this 
model is factorized: the many-particle scattering process is reduced to the
two-particle ones and the sets of two-momenta of incoming and outgoing 
particles are identical. It is well known since 1975 that the SG model is 
equivalent to the Massive Thirring model (MTM)  \cite{Col} with the action
\begin{equation}
{\cal S}_{MTM}=\int d^{2}x\left\{ i\overline{\psi }\gamma ^{\nu }\partial
_{\nu }\psi -M\ \overline{\psi }\psi -\frac{g}{2}\left( \overline{\psi }
\gamma ^{\nu }\psi \right) \left( \overline{\psi }\gamma _{\nu }\psi \right)
\right\} ,  \label{0.2}
\end{equation}
where $\overline{\psi }$, $\psi $ are two component Dirac spinors. 
This equivalence assumes the identification of the fundamental (anti) 
Fermions of the action (\ref{0.2}) with the (anti) solitons of sine-Gordon 
model and the following relations among parameters and currents, established by 
Coleman  \cite{Col}:
\begin{equation}
\frac{g}{\pi }=\frac{1}{2\beta ^{2}}-1;\ J^{\nu }\equiv \overline{\psi }
\gamma ^{\nu }\psi =-\frac{\beta }{2\pi }\epsilon ^{\nu \mu }\partial _{\mu
}\varphi  \label{0.3}
\end{equation}
 
More recently Al.Zamolodchikov has obtained an exact relation between the 
soliton mass $M$ and the parameter $\mu$ in the action (\ref{0.1}) 
\cite{AlZam} 
\begin{equation}
\mu =\frac{\Gamma \left( \beta ^{2}\right) }{\pi \Gamma \left( 1-\beta
^{2}\right) }\left[ \frac{M\sqrt{\pi }\Gamma \left( \frac{1+\xi }{2}\right) 
}{2\Gamma \left( \frac{\xi }{2}\right) }\right] ^{2-2\beta ^{2}},
\label{0.4}
\end{equation}
where 
\begin{equation}
\xi =\frac{\beta ^{2}}{1-\beta ^{2}}. \label{0.5}
\end{equation}

In this paper we consider  the Vacuum Expectation Values (VEV) of exponential 
fields in the sine-Gordon model
\begin{equation}
G_{a}=\left\langle \exp ia\varphi \left(0\right) \right\rangle, \label{0.6}
\end{equation}
where the exponential fields are normalized by the condition 
\begin{equation}
\left\langle e^{ia\varphi \left( x\right) }e^{-ia\varphi \left( y\right)
}\right\rangle _{SG}\rightarrow \left| x-y\right| ^{-4a^{2}}\quad as\ \left|
x-y\right| \rightarrow 0,  \label{0.7}
\end{equation}
which emphasizes that the UV limit of this theory is governed by the $c=1$ 
conformal free bosonic field.

For two special values of sine-Gordon coupling constant, namely for 
$\beta \rightarrow 0$ (semiclassical limit) and $\beta ^{2}=1/2$ (free 
fermion case), this function admits a direct calculation. 
The authors of \cite{LukZam} have used these special cases to guess the 
following expression for the expectation values (\ref{0.6}) for generic 
$\beta^2<1$ and $| Re(a)|<1/(2\beta)$
\begin{eqnarray}
\label{0.8}
&&G_{a}=\left( \frac{m\Gamma \left( \frac{1+\xi }{2}\right) 
\Gamma \left( \frac{2-\xi }{2}\right) }
{4\sqrt{\pi }}\right) ^{2a^{2}}\times  \\
&&\exp \left\{ \int\limits_{0}^{\infty }
\frac{dt}{t}\left[ \frac{\sinh^{2}
\left( 2a\beta t\right) }{2\sinh (\beta ^{2}t)
\sinh t\cosh \left( \left(1-\beta ^{2}\right) t\right) }
-2a^{2}e^{-2t}\right] \right\} .  \nonumber
\end{eqnarray} 

In order to support the formula (\ref{0.8}), some extra arguments, based 
on the reflection relations with Liouville reflection amplitude have 
been presented in the subsequent papers \cite{FatLukZamZam1}, 
\cite{FatLukZamZam2}, but there is no rigorous mathematical proof yet. 
The article \cite{P} provides another evidence supporting the 
Lukyanov-Zamolodchikov formula (\ref{0.8}), where its correctness has been 
checked in first order of MTM coupling constant $g$, by us of perturbation 
theory in angular quantization approach. 

In present paper we apply radial quantization to the same problem. 
The Hamiltonian of massive theories in the radial quantization approach 
has explicit time dependence \cite{Jakiw}. It appeared that this apparent 
difficulty can be overcome. We believe that such perturbative calculations 
substantially increase the confidence in reflection relations method as 
whole, which appears to be very powerful tool for investigation of 2d 
Conformal Field Theory (CFT) and IQFT \cite{Fat1}, \cite{Fat2}.

The paper is organized as follows. In section 2 we introduce the radial 
quantization of MTM. In section 3 we calculate  the VEV (\ref{0.6}) at free 
fermion point $g=0$. The calculation of VEV in the first order of perturbation
theory is presented in section 4. Here a special attention has been paid
to regularization procedure of the product of local fields at the  
coinciding points, which has some new features in comparison with the 
ordinary quantization in Cartesian coordinates. It appears that the 
Hankel-transform is a useful tool to carry out the calculations of 
section 4. The relevant mathematical details are presented in Appendix.

\vspace{0.5cm} \setcounter{equation}{0}

\section{Radial Quantization of the Massive Thirring Model}

\renewcommand{\theequation}{2.\arabic{equation}}

In two dimensional space it is convenient to use the following representation 
of Dirac matrices

\begin{equation}
\gamma ^0=\sigma _2=\left( 
\begin{array}{lr}
0 & -i \\ 
i & 0
\end{array}
\right) ,\ \gamma ^1=-i\sigma _{1}=-i\left( 
\begin{array}{ll}
0 & 1 \\ 
1 & 0
\end{array}
\right)  \label{1.1}
\end{equation}
and denote the components of Dirac spinors as

\begin{equation}
\psi \equiv \left( 
\begin{array}{l}
\psi _L \\ 
\psi _R
\end{array}
\right) ,\ \overline{\psi }\equiv \psi ^{\dagger }\gamma ^0.  \label{1.2}
\end{equation}
In this notations the action (\ref{0.7}) in Euclidean space takes the
factorized form

\begin{eqnarray}
{\cal A}_{MTM} &=&\int d^2z\left[ \psi _R^{\dagger }\partial \psi _R+\psi
_L^{\dagger }\overline{\partial }\psi _L-\frac{iM}2\left( \psi _L^{\dagger
}\psi _R-\psi _R^{\dagger }\psi _L\right) \right.  \nonumber \\
&&\ \qquad \qquad \qquad \qquad \qquad \qquad \qquad \quad \left. +g\psi
_L^{\dagger }\psi _L\psi _R^{\dagger }\psi _R\right] ,  \label{1.3}
\end{eqnarray}
where $z=x^2+ix^1$, $\overline{z}=x^2-ix^1$ are the complex coordinates 
on the Euclidean plane, $\partial \equiv \partial/\partial z$, 
$\overline{\partial }\equiv \partial /\partial \overline{z}$
and $d^2z \equiv 2dx^1dx^2$ is the volume element.

As we are interested in the VEV's of local fields $\left\langle
e^{ia\varphi \left( 0\right) }\right\rangle $, which have rotational
symmetry, it is natural to use the polar coordinates $\eta $, $\theta $
defined by

\begin{equation}
z= e^{\eta +i\theta };\quad \overline{z}= e^{\eta -i\theta }
\label{1.4}
\end{equation}
and interpret $\eta $, $\theta $ as Euclidean time and space coordinates
respectively.

Since the conformal dimensions of the Fermi fields $\psi_L$ and $\psi_R$ are
$(1/2,0)$ and $(0,1/2)$, they behave  under the conformal transformations 
(\ref{1.4}) as

\begin{equation}
\psi _L\left( z,\overline{z}\right) \rightarrow 
\left( \frac{\partial \xi }{\partial z}
\right) ^{\frac 12}\psi _L\left( \eta ,\theta \right) ;
\ \psi _R\left( z,\overline{z}\right) \rightarrow 
\left( \frac{\partial \overline{\xi }}
{\partial \overline{z}}\right) ^{\frac 12}\psi _R\left( \eta ,\theta \right) ,
\label{1.5}
\end{equation}

where $\xi =\eta +i\theta ,\overline{\xi }=\eta -i\theta$. The same 
transformation lows hold for the fields $\psi ^{\dagger}_{L,R}$.

Thus in $\left( \eta ,\theta \right) $ coordinates the action 
(\ref{1.3}) becomes

\begin{eqnarray}
{\cal A}_{MTM} &=&\int\limits_0^{2\pi }d\theta 
\int\limits_{-\infty}^\infty d\eta \ \left[ i\psi _L^{\dagger }
\left( \partial _\theta -i\partial _\eta \right) \psi _L-i\psi _R^{\dagger }
\left( \partial _\theta +i\partial _\eta \right) \psi _R-\right.  \nonumber \\
\quad \qquad \qquad &&\qquad \quad \left. iMe^\eta \left( \psi _L^{\dagger
}\psi _R-\psi _R^{\dagger }\psi _L\right) +2g\psi _L^{\dagger }
\psi _L\psi _R^{\dagger }\psi _R\right] .  \label{1.6}
\end{eqnarray}

Treating the radial coordinate $\eta $ as a "time", we get the Hamiltonian
\begin{eqnarray}
H &=&\int\limits_0^{2\pi }d\theta \left[ \psi _L^{\dagger }i\partial
_\theta \psi _L-\psi _R^{\dagger }i\partial _\theta \psi _R-iMe^\eta \left(
\psi _L^{\dagger }\psi _R-\psi _R^{\dagger }\psi _L\right) \right. 
\nonumber \\
&&\qquad \qquad \qquad \qquad \qquad \qquad \qquad \quad \left. +2g\psi
_L^{\dagger }\psi _L\psi _R^{\dagger }\psi _R\right] .  \label{1.7}
\end{eqnarray}

The usual canonical quantization scheme will bring us to the standard 
''equal time'' anti-commutation relations

\begin{equation}
\left\{ \psi _L\left( \theta \right) ,\psi _L^{\dagger }\left( \theta
^{\prime }\right) \right\} =\delta \left( \theta -\theta ^{\prime }\right)
,\quad \left\{ \psi _R\left( \theta \right) ,\psi _R^{\dagger }\left( \theta
^{\prime }\right) \right\} =\delta \left( \theta -\theta ^{\prime }\right) .
\label{1.8}
\end{equation}

As usual, in order to develop perturbation theory one first has to solve
the problem with unperturbed Hamiltonian (i.e. to ignore the last quartic 
term in (\ref{1.7})). We found it easier to handle this problem in 
Schr\"odinger picture, instead of more conventional in QFT Heisenberg 
or Interaction pictures. Thus our field operators $\psi_{L,R}$ will 
not depend on ``time'' $\eta$ and the state vectors will evolve according
to the Schr\"odiger equation. Let us define the creation, annihilation operators
$c_k^{\dagger}$, $d_k^{\dagger}$, $c_k$, $d_k$ through the Fourier mode 
decompositions

\begin{eqnarray}
\label{1.9}
\psi _L\left( \theta \right) &=&\frac 1{\sqrt{2\pi }}\sum\limits_{k\in 
{\cal N-}\frac 12}\left( c_k \ e^{-ik\theta }+d_k^{\dagger } \ 
e^{ik\theta }\right) \nonumber \\
\psi _R\left( \theta \right) &=&\frac 1{\sqrt{2\pi }}\sum\limits_{k\in 
{\cal N-}\frac 12}\left( c_{-k} \ e^{ik\theta }+d_{-k}^{\dagger } \ 
e^{-ik\theta } \right)  \nonumber \\
\psi _L^{\dagger }\left( \theta \right) &=&\frac 1{\sqrt{2\pi }}
\sum\limits_{k\in {\cal N-}\frac 12}\left( d_k \ e^{-ik\theta }+c_k^{\dagger} 
\ e^{ik\theta }\right)  \nonumber \\
\psi _R^{\dagger }\left( \theta \right) &=&\frac 1{\sqrt{2\pi }}
\sum\limits_{k\in {\cal N-}\frac 12}\left( d_{-k} \ e^{ik\theta }+
c_{-k}^{\dagger} \ e^{-ik\theta }\right)
\end{eqnarray}

where all sums are taken over all positive half-integers (${\cal N}$ is 
the set of positive integers).  

As a consequence of equations (\ref{1.8}) and (\ref{1.9}), one can easily 
get the following anti-commutation relations for the operators 
$c_k,d_k,c_k^{\dagger },d_k^{\dagger }$:

\begin{eqnarray}
&&\left\{ c_k,c_l\right\} =\left\{ c_k^{\dagger },c_l^{\dagger }\right\}
=\left\{ d_k,d_l\right\} =\left\{ d_k^{\dagger },d_l^{\dagger }\right\} =0, 
\nonumber \\ 
&&\left\{ c_k,c_l^{\dagger }\right\} =\delta _{k,l},
\quad \left\{d_k,d_l^{\dagger }\right\} 
=\delta _{k,l},\quad k,l\in {\cal Z}-\frac 12 , 
\label{1.10}
\end{eqnarray}

where ${\cal Z}$ is the set of integers.
As usual the Fock space (let us denote it ${\cal H}$) has the following basic 
vectors

\begin{equation}
\prod\limits_{k\in {\cal Z}-\frac 12}\left(c_k^{\dagger}\right)^{n_k}
\left(d_k^{\dagger}\right)^{\tilde n_k}|0\rangle,
\label{basevector}
\end{equation}

where $n_k \in \{0,1 \}$ ($\tilde n_k \in \{ 0,1 \}$) are the occupation
numbers of ``particles'' created by the operators $c_k^{\dagger}$ 
($d_k^{\dagger}$) out of bare vacuum $|0 \rangle $, which by definition
satisfies the conditions

\begin{equation}
\label{barevac}
c_k |0 \rangle = d_k |0 \rangle = 0; \quad k\in {\cal Z}-\frac 12.
\end{equation}

In terms of the creation, annihilation operators is the Hamiltonian 
(\ref{1.7}) acquires the form $H=H_0+H_{int}$, where the quadratic part 
$ H_0$ is equal to

\begin{eqnarray}
\label{ham}
&&H_0=\sum\limits_{k\in {\cal N}-\frac 12}\left[ k\left( c_k^{\dagger}
c_k-d_kd_k^{\dagger }+c_{-k}^{\dagger }c_{-k}-d_{-k}d_{-k}^{\dagger}
\right)\right.\\
&& \hspace{3cm}-iMe^\eta \left . \left( c_k^{\dagger }d_{-k}^{\dagger}
-d_{-k}c_k+d_kc_{-k}-c_{-k}^{\dagger }d_k^{\dagger }\right) \right]\nonumber
\end{eqnarray}

The evolution of arbitrary state $|s\rangle $ along Euclidean time 
$\eta $ caused by the Hamiltonian $H_0$ is given by the Schr\"odinger
equation

\begin{equation}
\label{Sch}
-r\frac{\partial }{\partial r }|s,r\rangle = H_{0}|s,r\rangle .
\end{equation}

Here and henceforth we prefer to use $r\equiv Me^{\eta}$ rather than 
$\eta $.
To find the general solution to the Schr\"odinger equation (\ref{Sch}) 
let us denote, that the Hamiltonian $H_0$ has a factorized form

\begin{equation}
\label{Hamfac} 
H_0=\sum\limits_{k\in {\cal N}-\frac 12} \left(H_k^{(1)}+H_k^{(2)}\right),
\end{equation}

where the operator $H_k^{(1)}$ ($H_k^{(2)}$) includes only $c_k,d_{-k},
c_k^{\dagger},d_{-k}^{\dagger}$ ($d_k,c_{-k},d_k^{\dagger},c_{-k}^{\dagger}$).
This makes convenient to represent the Fock space ${\cal H}$ as an infinite
tensor product

\begin{equation}
\label{spacefac}
{\cal H}=\otimes_{k\in{\cal N}-1/2}\left( {\cal H}_k^{(1)}
\otimes {\cal H}_k^{(2)}\right),
\end{equation}

where ${\cal H}_k^{(1)}$ and ${\cal H}_k^{(2)}$ are four dimensional vector
spaces with base vectors

\begin{eqnarray}
\label{basevec1}
|0_k^{(1)} \rangle, \quad c_k^{\dagger}d_{-k}^{\dagger}|0_k^{(1)} \rangle ,
\quad \mbox{ (even sector)} \nonumber \\
c_k^{\dagger}|0_k^{(1)} \rangle , \quad d_{-k}^{\dagger}|0_k^{(1)} \rangle ,
\quad \mbox{ (odd sector)}
\end{eqnarray}

and

\begin{eqnarray}
\label{basevec2}
|0_k^{(2)} \rangle , \quad d_k^{\dagger}c_{-k}^{\dagger}|0_k^{(2)} \rangle ,
\quad \mbox{ (even sector)} \nonumber \\
d_k^{\dagger}|0_k^{(2)} \rangle , \quad c_{-k}^{\dagger}|0_k^{(2)} \rangle , 
\quad \mbox{ (odd sector)}
\end{eqnarray}

respectively. The vectors $|0_k^{(1)} \rangle $ and $|0_k^{(2)} \rangle $ are 
defined by the conditions

\begin{eqnarray}
\label{vacdef}
c_k|0_k^{(1)} \rangle =d_{-k}|0_k^{(1)} \rangle =0, \nonumber \\
d_k|0_k^{(2)} \rangle =c_{-k}|0_k^{(2)} \rangle =0,
\end{eqnarray}

for any $k\in {\cal N}-1/2$. Note that the bare vacuum introduced earlier 
(see (\ref{barevac}) ) is equal to

\begin{equation}
\label{vacvacsrel}
|0 \rangle =\otimes_{k \in {\cal N}-\frac 12}\left[ |0_k^{(1)} \rangle 
\otimes |0_k^{(2)} \rangle \right].
\end{equation}

The operator $H_k^{(1)}$ ($H_k^{(2)}$) nontrivially acts only on the
factor ${\cal H}_k^{(1)}$ (${\cal H}_k^{(2)}$) of the full Fock space 
${\cal H}$ (\ref{spacefac}), hence we have reduced the initial QFT problem of 
infinitely many degrees of freedom to the simple quantum mechanical one, with
four dimensional Hilbert space. A further simplification provides the 
observation, that the reduced Hamiltonians $H_k^{(1)}$, $H_k^{(2)}$
don't mix even and odd sectors (see (\ref{basevec1}), (\ref{basevec2})). 
The resulting Schr\"odinger equations in this reduced spaces take the
form

\begin{eqnarray}
\label{redScheven}
-r\frac{\partial}{\partial r}\left[\alpha_k (r)+\beta_k (r)
c_k^{\dagger}d_{-k}^{\dagger}\right]|0_k^{(1)} \rangle =
H_k^{(1)}\left[\alpha_k (r)+\beta_k (r)
c_k^{\dagger}d_{-k}^{\dagger}\right]|0_k^{(1)} \rangle = \nonumber \\
\left[-k\alpha_k (r)+ir\beta_k (r)+(k\beta_k (r)-ir\alpha_k (r))
c_k^{\dagger}d_{-k}^{\dagger} \right]|0_k^{(1)} \rangle ,
\end{eqnarray}

and

\begin{equation}
\label{redSchodd}
-r\frac{\partial}{\partial r}\left[\gamma_k (r)c_k^{\dagger}+
\delta_k (r) d_{-k}^{\dagger} \right] |0_k^{(1)} \rangle = 
H_k^{(1)}\left[\gamma_k (r)c_k^{\dagger}+\delta_k (r) d_{-k}^{\dagger}
\right]|0_k^{(1)} \rangle =0.
\end{equation}

Evidently, to obtain the equations for the another sector with Hamiltonian
$H_k^{(2)}$, one simply has to change the upper indices $(1)$ into $(2)$ and
make the substitutions $c_k \rightarrow d_k$ and $d_{-k} \rightarrow 
c_{-k}$. Thus, the differential equations for the unknown functions 
$\alpha_k(r)$, $\beta_k(r)$, $\gamma_k(r)$, $\delta_k(r)$ in both cases remain
the same:

\begin{eqnarray}
\label{difeq1}
&& \left(\frac{\partial}{\partial r}-\frac{k}{r} \right)\alpha_k(r)=
-i\beta_k(r), \nonumber \\
&& \left(\frac{\partial}{\partial r}+\frac{k}{r} \right)\beta_k(r)=
i\alpha_k(r) \nonumber \\
&&\frac{\partial}{\partial r}\gamma_k(r)=
\frac{\partial}{\partial r}\delta_k(r)=0.
\end{eqnarray}

The second pare of these equations show, that in fact $\gamma_k$ and 
$\delta_k$ are constants, while the first pare reduces to the modified 
Bessel differential equation with general solution
  
\begin{eqnarray}
&&\alpha _k\left( r\right)=r^{\frac 12}\left( a_kI_{k-\frac 12}
\left(r\right) +b_kK_{k-\frac 12}\left( r\right) \right) ,  \nonumber \\
&&\beta _k\left( r\right)=ir^{\frac 12}\left( a_kI_{k+\frac 12}
\left(r\right) -b_kK_{k+\frac 12}\left( r\right) \right) .  
\label{1.13}
\end{eqnarray}

One should fix the constants $a_k,b_k,\gamma_k $ and $\delta_k$ imposing
initial conditions at the arbitrary ``time'' $r_0$. For the further 
application let us write down explicit expressions with specified constants
for two basic cases \\
a) when the initial state coincides with $|0_k^{(1)}\rangle $ or 
$|0_k^{(2)}\rangle $ :
\begin{eqnarray}
\label{inconda}
&&\alpha _k\left( r\right)=\sqrt{rr_0}\left( K_{k+\frac 12}\left(r_0\right)
I_{k-\frac 12}\left(r\right) +I_{k+\frac 12}\left(r_0\right)
K_{k-\frac 12}\left( r\right) \right) ,  \nonumber \\
&&\beta _k\left( r\right)=i\sqrt{rr_0}\left( K_{k+\frac 12}\left(r_0\right)
I_{k+\frac 12}\left(r\right) -I_{k+\frac 12}\left(r_0\right)
K_{k+\frac 12}\left( r\right) \right) ,
\end{eqnarray}
b) when the initial state coincides with 
$ic_k^{(\dagger)}d_{-k}^{(\dagger)}|0_k^{(1)}\rangle $ or 
$id_k^{(\dagger)}c_{-k}^{(\dagger)}|0_k^{(2)}\rangle $
\begin{eqnarray}
\label{incondb}
&&\alpha _k\left( r\right)=\sqrt{rr_0}\left( K_{k-\frac 12}\left(r_0\right)
I_{k-\frac 12}\left(r\right) -I_{k-\frac 12}\left(r_0\right)
K_{k-\frac 12}\left( r\right) \right) ,  \nonumber \\
&&\beta _k\left( r\right)=i\sqrt{rr_0}\left( K_{k-\frac 12}\left(r_0\right)
I_{k+\frac 12}\left(r\right) +I_{k-\frac 12}\left(r_0\right)
K_{k+\frac 12}\left( r\right) \right) .
\end{eqnarray}
During the proof of the formulae (\ref{inconda})(\ref{incondb}) we have 
the standard Wronskian identity for the modified Bessel functions \cite{BatErd}
\begin{equation}
I_{k-\frac 12}\left( r\right) K_{k+\frac 12}\left( r\right) +
I_{k+\frac 12}\left( r\right) K_{k-\frac 12}\left( r\right) =\frac 1r.  
\label{Wronskian}
\end{equation}

It is interesting to note, that due to explicit time dependence of the
Hamiltonian, the system being initially at the ground state of that 
particular moment, after finite time of evolution will find himself in an 
excited state. Nevertheless, long time evolution of any state with
non vanishing overlap with the ground state of the initial time, 
eventually approaches to the ground state of the infinite future  

\begin{equation}
\label{futurevac}
|\infty \rangle \equiv \prod\limits_{k\in {\cal N}-\frac 12}
\left[ \frac 12\left( 1+ic_k^{\dagger}d_{-k}^{\dagger}\right) 
\left( 1+id_k^{\dagger}c_{-k}^{\dagger}\right) \right]|0\rangle .
\end{equation}

Evidently, at the small $r$'s (far past), the ground state approaches to
the bare vacuum $|0\rangle $. In particular, if $r \gg 1$ and 
$r_0 \ll 1$ 
(\ref{inconda}) gives

\begin{eqnarray}
\label{asymp}
\alpha_k \left(r\right)\rightarrow \left(\frac{2}{r_0}\right)^k 
\frac{e^r}{\sqrt{4\pi}}\Gamma \left(k+\frac 12 \right), \nonumber \\
\beta_k \left(r\right) \rightarrow \left(\frac{2}{r_0}\right)^k 
\frac{e^r}{\sqrt{4\pi}}\Gamma \left(k+\frac 12 \right) .
\end{eqnarray}

\section{The VEV's of the Exponential Fields in \\ Free Fermion Case }

\renewcommand{\theequation}{3.\arabic{equation}}

The VEV (\ref{0.6})
\begin{equation}
\label{3.1}
G_a =\left\langle e^{ia\varphi } \left( 0\right)\right\rangle =\frac{{\int }
{\cal D}\varphi e^{ia\varphi }e^{-
{\cal S}_{SG}\left( \varphi \right) }}{{\int }{\cal D}\varphi
e^{-{\cal S}_{SG}\left( \varphi \right) }},
\end{equation}
with ${\cal S}_{SG}$ being  the action (\ref{0.1}), can be expressed 
alternatively in terms of the  appropriately regularized (see below) Euclidean 
functional integral over the Dirac fermions 
\begin{equation}
\label{3.2}
G(a)=\frac{\int\limits_{{\cal F}_a}\left[ {\cal D}\psi {\cal D}
\overline{\psi }\right] e^{-{\cal A}_{MTM}}}{\int\limits_{{\cal F}_0}
\left[ {\cal D}\psi {\cal D}
\overline{\psi }\right] e^{-{\cal A}_{MTM}}}
\end{equation}
where ${\cal A}_{MTM}$ is the Euclidean  action (\ref{1.6}).
The functional integral in the numerator of (\ref{3.2}) is taken over 
the space ${\cal F}_a$ of the twisted field configurations 
$\psi \left( z,\overline{z}\right)$ and 
$\overline{\psi }\left( z,\overline{z}\right) $, which acquire the phase

\begin{equation}
\psi \left( z,\overline{z}\right)\rightarrow e^{i\frac{2\pi a}\beta }\psi 
\left( z,\overline{z}\right),\qquad \overline{\psi }\left( z,\overline{z}
\right) \rightarrow  e^{-i\frac{2\pi a}\beta }
\overline{\psi }\left( z,\overline{z}\right),
\label{3.3}
\end{equation}
when continued analytically around the point $z=0$ in anti-clockwise direction
\cite{LukZam}. This is due to the non-trivial monodromy of Dirac fields 
with respect to the exponential fields $\exp ia\varphi \left( 0\right) $ .
It is easy to see, that to ensure such twisted boundary conditions on 
Dirac fields, one has to shift Fourier mode indices as follows
\begin{eqnarray}
\label{indshift}
k\rightarrow k-\frac{a}{\beta},\quad \mbox{ in sector 1 (i.e. in $c_k, d_{k}$ 
sector)} \nonumber \\
k\rightarrow k+\frac{a}{\beta},\quad \mbox{ in sector 2 (i.e. in $d_k, c_{k}$ 
sector)}.
\end{eqnarray}
For example, the Fourier decomposition of the field $\psi_L (\theta)$ takes 
the form (cf. with the first line of (\ref{1.9}))
\begin{equation}
\label{shiftexample}
\psi _L\left( \theta \right) =\frac 1{\sqrt{2\pi }}\sum\limits_{k\in 
{\cal N-}\frac 12}\left( c_{k-\alpha} \ e^{-i(k-\alpha)\theta }+
d_{k-\alpha}^{\dagger } 
\ e^{i(k-\alpha)\theta }\right),
\end{equation}
where 
\begin{equation}
\label{defalpha}
\alpha \equiv \frac{a}{\beta}.
\end{equation}
With such shifts, all the results of the previous sector remain valid since
we have never used the arithmetical properties of the Fourier mode indices.

In radial Hamiltonian formalism the regularized version of the functional
integral (\ref{3.2}) may be represented as
\begin{equation}
\label{3.5}
G(a,r_0) =\frac{\left\langle \infty \right| S(r,r_0)\left|0  
\right\rangle _{a}}
{\left\langle \infty \right| S(r,r_0)\left|0 \right\rangle},
\end{equation}
where the matrix element of the evolution operator $S(r,r_0)$ in the 
numerator is taken in twisted sector (this is indicated by the lower index 
$a$). To regularize the expression, in (\ref{3.5}) we have assumed, that the
evolution begins at some small $r_0$. A simple Conformal Field Theory 
consideration\footnote{In the limit $r\rightarrow 0$ the action (\ref{1.6})
describes the Massless Thiring Model, which is well known to be conformal
invariant.}, which takes into account the fact, that the conformal dimension 
of the field $e^{ia\varphi }$ is $a^2$, leads to
\begin{equation}
G_0(a)=\lim\limits_{r_0 \rightarrow 0}\left( r_0 \right)^{-2a^2}G_0(a,r_0).
\label{r0dependence}
\end{equation}
In general case it is not known yet how to calculate the functional 
integral (\ref{3.2}) or the matrix elements in (\ref{3.5}) exactly. 
Below we'll evaluate (\ref{3.5}) at the free fermion 
point $g=0$. As in this case we already know the time evolution of every 
constituent of the vacuum $|0\rangle $ (see eq. (\ref{vacvacsrel})) from the
previous section, it is not difficult to pick up all the necessary factors
from (\ref{asymp}) with appropriate shifts of Fourier mode indices and obtain
\begin{eqnarray}
&&\hspace {-1cm}G_0(a,r_0)=\\ 
&&\hspace {-0.4cm}\lim\limits_{r\rightarrow \infty }
\frac{\prod\limits_{k \in {\cal N}-\frac 12} 
\left( \frac 2{r_0}\right) ^{k+\alpha }\hspace {-0.2cm}
\frac{e^r}{\sqrt{4\pi }}\Gamma 
\left( k+\alpha +\frac 12\right) 
\prod\limits_{l\in {\cal N}-\frac 12}
\left( \frac 2{r_0}\right) ^{l-\alpha}
\hspace {-0.2cm}\frac{e^r}{\sqrt{4\pi }}\Gamma \left( l-\alpha +
\frac 12\right)}
{\prod\limits_{k\in {\cal N}-\frac 12}
\left( \frac 2{r_0}\right) ^{k}\hspace {-0.2cm}\frac{e^r}{\sqrt{4\pi }}
\Gamma \left( k+\frac 12\right) 
\prod\limits_{l\in {\cal N}-\frac 12}
\left( \frac 2{r_0}\right) ^{l}
\hspace {-0.2cm}\frac{e^r}{\sqrt{4\pi }}
\Gamma \left( l+\frac 12\right)}, \nonumber
\label{bigprod}
\end{eqnarray}
where we have added a subscript $0$ to $ G $ in order to 
emphasize, that the free Fermion case $g=0$ is considered.
We have to be careful, when evaluating infinite products in (\ref{bigprod}) 
and treat the ill defined sums like $\sum\limits_{i=0}^{\infty}(i\pm a)$
by means of Riemann $\zeta $-function regularization. Let us remind that 
\begin{equation}
\label{zeta}
\zeta \left( z,a\right) =\sum\limits_{i=0}^{\infty }
\frac 1{\left( i+a\right) ^z}
\end{equation}
and
\begin{equation}
\label{zetaprop}
\zeta \left( -1,a\right) +\zeta \left( -1,-a\right) -2\zeta
\left( -1,0\right) =-a^2
\end{equation}
To carry out the remaining infinite products of $\Gamma$-functions 
(also divergent, if treated literally), it is convenient to use the integral
representation 
\begin{equation}
\ln \Gamma \left( \nu\right) =
\int\limits_{0}^{\infty }
\left[ \frac{e^{- \nu t}-e^{-t}}{1-e^{-t}}+\left( \nu-1\right)
e^{-t}\right] \frac{dt}t . 
\label{3.9}
\end{equation}  
As a result we obtain simple geometric progressions, coming from the 
first term of the eq. (\ref{3.9}) and contributions, coming from the 
second term, which can be easily handled applying $\zeta$-function 
regularization once more. The final expression has the form
\begin{equation}
G_0(a, r_0) =\left( \frac{r_0}2\right) ^{\alpha^2}\exp 
\int\limits_{0}^{\infty }\left[ \frac{\sinh ^2(\alpha t)}{\sinh ^2t}
-\alpha^2e^{-2t}\right] \frac{dt}t  
\label{3.11}
\end{equation}
or, taking into account equations (\ref{r0dependence}) and (\ref{defalpha}) 
with the free Fermion point value 
$\beta=1/\sqrt 2$
\begin{equation}
G_0(a) =\left( \frac{M}2\right) ^{2a^2}\exp \int\limits_{0}^{\infty }
\left[ \frac{\sinh ^2(\sqrt2 at)}{\sinh ^2t}-2a^2e^{-2t}\right] \frac{dt}t.  
\label{3.13}
\end{equation}
This is in full agreement with the result, obtained by S.Lukyanov and 
A.Zamolodchikov in \cite{LukZam}, using angular quantization technic.

\setcounter{equation}{0}

\section{VEV of the Exponential Field in the First Order of Perturbation
Theory}

\renewcommand{\theequation}{4.\arabic{equation}} \vspace{0.25cm}

In this section we calculate the VEV (\ref{0.6}) in first order of the 
MTM's coupling constant $g$. 
The perturbation is given by the last term of the Hamiltonian (\ref{1.7}):
\begin{equation}
{\cal H}_{int}=2g\int\limits_{0}^{2\pi }N\left( \Psi
_L^{+}\Psi _L\Psi _R^{+}\Psi _R\right) d\theta ,  \label{4.1}
\end{equation}
where we denoted by $N(\dots)$ an appropriately regularized product
of local fields at coinciding point. One has to chose such a regularization, 
which will not break the translational invariance of the theory if it
is transformed back to the initial Euclidean coordinates $x^1$, $x^2$. The 
conventional normal ordering with respect to creation-annihilation 
operators fails to satisfy this condition, because of the non-trivial 
time dependence of physical vacuum. Instead, we should suppress all the 
contractions among fields inside the correct normal ordering symbol $N(...)$ 
\begin{eqnarray}
&&N\left( \psi _L^{\dagger }\psi _L\psi _R^{\dagger }
\psi _R\right) = 
\psi_L^{\dagger }\psi _L\psi _R^{\dagger }\psi _R-
\left\langle \psi _L^{\dagger}\psi _L\right\rangle _{0}
\psi _R^{\dagger }\psi _R-
\left\langle \psi_R^{\dagger }\psi _R\right\rangle _{0}
\psi _L^{\dagger }\psi _L+  \nonumber\\
&&\left\langle \psi _L^{\dagger }\psi _R\right\rangle _{0}
\psi _R^{\dagger}\psi _L+
\left\langle \psi _R^{\dagger }\psi _L\right\rangle _{0}
\psi_L^{\dagger }\psi _R+ 
\left\langle \psi _L^{\dagger }\psi _L\right\rangle _{0}
\left\langle\psi _R^{\dagger }\psi _R\right\rangle _{0}-
\left\langle \psi _L^{\dagger}\psi _R\right\rangle _{0}
\left\langle \psi _R^{\dagger }\psi_L\right\rangle _{0} \nonumber \\
&&=:\psi _L^{\dagger }\psi _L\psi _R^{\dagger }\psi _R:+
\left\langle :\psi_L^{\dagger }\psi _R:\right\rangle _{0}
:\psi _R^{\dagger }\psi_L:+\left\langle 
:\psi _R^{\dagger }\psi _L:\right\rangle _{0}:
\psi_L^{\dagger }\psi _R: - \nonumber \\
&& \left\langle :\psi_L^{\dagger }\psi _L:
\right\rangle _{0}:\psi _R^{\dagger }\psi_R: 
-\left\langle :\psi _R^{\dagger }\psi _R:\right\rangle _{0}
:\psi_L^{\dagger }\psi _L:-
\left\langle :\psi _L^{\dagger }\psi _R:\right\rangle _{0}
\left\langle:\psi _R^{\dagger }\psi _L:\right\rangle _{0}+ \nonumber \\
&&\left\langle :\psi _L^{\dagger }\psi _L:\right\rangle _{0}
\left\langle:\psi _R^{\dagger }\psi _R:\right\rangle _{0},  
\label{4.2}
\end{eqnarray}
where the vacuum expectation value of any operator $X$ is defined by 
\begin{equation}
\label{4.2'}
\langle X \rangle_0 \equiv \frac{\langle \infty |S(R,r)XS(r,r_0|0\rangle)}
{\langle \infty |S(R,r_0) |0\rangle} 
\end{equation}
with all matrix elements taken in untwisted sector. The first part of
eq. (\ref{4.2}) could be understood for example as a zero distance limit
of corresponding point-split expression.  
In (\ref{4.2'}) a small initial time $r_0$ and a large final time 
$R_0$ are introduced in order to keep intermediate expressions finite. 
$R$ and $r_0$ eventually will be sent to $0$ and $\infty$ correspondingly. 
Note also the explicit $r$ dependence of (\ref{4.2'}) and hence of 
(\ref{4.2}), which reflects the inhomogeneity of ``time'' in our scheme
of quantization.
The standard time dependent perturbation theory in first order of the
coupling constant $g$ gives   
\begin{equation}
G\left(a,r_0\right)=\lim\limits_{R\rightarrow \infty}\left(
\left\langle \infty|S\left(R,r_0\right)|0\right\rangle_a +
\int \limits_{r_0}^{R}\left\langle \infty|S\left( R ,r\right) 
H_{int}S\left( r,r_0\right)|0 \right\rangle _{a}\frac{dr}{r}\right) ,  
\label{4.4}
\end{equation} 
As we already have obtained explicit expressions for time evolution of
states from various sectors of Fock space in section 2, it is not 
difficult to calculate the matrix element under the integral in
eq. (\ref{4.4}) 
\begin{eqnarray}
&&\hspace{-1cm}G\left(a,r_0\right)=
\lim\limits_{R\rightarrow \infty} \left\langle \infty 
\left|S\left(R,r_0\right) \right|0\right\rangle_a 
\left\{ 1+\right. \nonumber \\
&&\frac g\pi\sum\limits_{k,l=0}^{\infty} \int\limits_{r_0}^{R}
rdr\left [2I_{k+1-a}K_{k-a}I_{l+1+a}K_{l+a}- 
I_{k+1-a}K_{k-a}I_{l+1-a}K_{l-a}-\right. \nonumber \\
&&I_{k+1+a}K_{k+a}I_{l+1+a}K_{l+a}-I_{k-a}K_{k-a}I_{l-a+1}K_{l-a+1}-
I_{k+a}K_{k+a}I_{l+a+1}K_{l+a+1}- \nonumber \\
&&I_{k+1-a}K_{k+1-a}I_{l+1+a}K_{l+1+a}- 
I_{k+a}K_{k+a}I_{l-a}K_{l-a}+I_{k+1}K_{k+1}I_{l-a+1}K_{l-a+1}+\nonumber \\
&&I_kK_kI_{l-a+1}K_{l-a+1}+I_kK_kI_{l+a}K_{l+a}+I_kK_kI_{l+a+1}K_{l+a+1}+
I_kK_kI_{l-a}K_{l-a}+ \nonumber \\
&&I_{k+1}K_{k+1}I_{l+a+1}K_{l+a+1}+I_{k+1}K_{k+1}I_{l-a}K_{l-a}+
I_{k+1}K_{k+1}I_{l+a}K_{l+a}- \nonumber \\
&&\left.\left. I_{k+1}K_{k+1}I_{l+1}K_{l+1}-I_{k+1}K_{k+1}I_lK_l-
I_kK_kI_{l+1}K_{l+1}-I_kK_kI_lK_l\right]\right\},  
\label{4.5}
\end{eqnarray}
with the  pre factor $\langle \infty|S(R,r_0)|0\rangle_a$ given by eq. 
(\ref{3.11}) (in (\ref{3.11}) we have to insert 
$\beta =\frac{1}{\sqrt2}(1-\frac{g}{2\pi}+o(g) )$ and expand the resulting 
expression over $g$ up to linear term). The remaining calculation of 
integrals over the quartic Bessel functions is presented in the appendix. 
Using these results we obtain
\begin{eqnarray}
\label{intermed}
&&G\left(a,r_0\right)=\left(\frac{r_0}{2}\right)^{\alpha^2} 
\exp \left\{\int\limits_{0}^{\infty }
\left( \frac{\sinh ^{2}(\alpha t)}{\sinh ^{2}t}-\alpha^{2}e^{-2t}\right) 
\frac{dt}{t}\right\} \times  \\
&&\left\{1+\frac{g}{2\pi}\left[2\alpha^2 \log \frac{r_0}{2}+
\int\limits_{0}^{\infty }
\left(\frac{\alpha \sinh(2\alpha t)}{2\sinh ^{2}t}-
\frac{\alpha^2}{t}e^{-2t}\right) dt\right] +O(g^2) \right\}\times \nonumber \\
&&\left\{1+\frac{g}{\pi}\left[\int\limits_0^{\infty}
\sum\limits_{k,l=0}^{\infty}\left(\frac{8\cosh^2 t \sinh^2 \alpha t-
4\sinh^2 2\alpha t}{\sinh 2t} \ e^{-2(k+l+1)t} \right)dt\right]+
O(g^2)\right\}. \nonumber
\end{eqnarray} 
Now performing summation over $k$ and $l$ in the third line of the eq. 
(\ref{intermed}) we obtain logarithmically diverging at $t=0$ integral. 
In fact, the same problem we have faced when carrying out calculations 
exactly at the free-fermion point. Indeed the the product in (\ref{bigprod}) 
diverges for large $k$, $l$, but we overcame this difficulty using 
$\zeta$-function regularization inside the integral representation of 
$\Gamma$-function (\ref{3.9}). Here we'll not care of a similar appropriate 
regularization. Rather, noticing that various regularization scheme will 
differ from each other by a term $\sim a^2$, and that the coefficient of 
$-1/2a^2$ in the expansion of $\langle e^{ia\phi}\rangle$ is just the VEV 
$\langle \phi^2\rangle$, which has been calculated in \cite{LukZam} using 
standard Feynman diagram technic with result (below $\gamma =0.577216...$ 
is the Euler constant) 
\begin{equation}
\label{vevphi2}
\langle \phi^2(0)\rangle =-4(1+\gamma +\log (M/2))+
\frac{g}{\pi}(7\zeta(3)-2)+O(g^2), 
\end{equation}    
we simply cut the above mentioned integral over $t$ on the lower bound and
equate the undefined coefficient of $-1/2a^2$ to the one predicted by 
the eq.(\ref{vevphi2}). The final result is
\begin{eqnarray}
&&\left\langle e^{ia\phi(0)}\right\rangle =
\left( \frac{M}{2}\right) ^{\alpha^{2}}
\exp \left\{\int\limits_{0}^{\infty }
\left( \frac{\sinh ^{2}(\alpha t)}{\sinh ^{2}t}-\alpha^{2}e^{-2t}\right) 
\frac{dt}{t}\right\} \times\\
&& \left\{1+\frac{g}{\pi }
\left[\int\limits_{0}^{\infty }
\left(\frac{\alpha \sinh(2\alpha t)}{2\sinh ^{2}t}-
\frac{\sinh ^{2}(\alpha t)}{\sinh ^{3}t}
\right) dt- 2\alpha^2\log 2\right] +O(g^2) \right\} \nonumber,  
\label{4.9}
\end{eqnarray}
which is in complete agreement with the Lukyanov-Zamolodchikov formula 
(\ref{0.8}). 

\vspace{1cm}

\section*{Acknowledgments} 

We are grateful to A.Sedrakyan for useful discussions. R.P. thanks 
theory division of LAPP and especially P.Sorba for warm 
hospitality and discussions. The work of V.M. and R.P. 
was partially supported by INTAS grant 96-690. 

\vspace{1cm} \setcounter{equation}{0} 
\section*{ Appendix } 
\renewcommand{\theequation}{A.\arabic{equation}} \vspace{0.25cm}

It appears that the Hankel-transforms are appropriate tools allowing us to
perform the integration over $r$ in (\ref{4.5}). Roughly speaking, in polar 
coordinates the Hankel-transforms play the same role, as the ordinary 
Fourier-transforms in the Cartesian one.

Let us briefly recall the main formulae concerning to the Hankel-transforms
(for details see \cite{BatErd} and references therein). The $\nu $-th order
( $\nu >-1$ ) direct and inverse Hankel-transforms of the function $f(x)$
defined on $(0,\infty )$ are given by 
\begin{equation}
f\left( x\right) =\int\limits_{0}^{\infty }J_{\nu }\left( sx\right) 
\widetilde{f_{\nu }}\left( s\right) sds,  \label{A1}
\end{equation}
\begin{equation}
\widetilde{f_{\nu }}\left( s\right) =\int\limits_{0}^{\infty }J_{\nu }\left(
sx\right) f\left( x\right) xdx,  \label{A2}
\end{equation}
where $J_{\nu }$ is the Bessel function. In complete analogy with the case
of Fourier-transform, it follows from (\ref{A1}), (\ref{A2}), that the
``scalar product'' of any two functions $f(x)$, $g(x)$ coincides
with that of their images
\footnote
{Since the functions we are dealing with are regular in the interval 
$(0,\infty )$, the only thing one has to care of is the convergence of
integrals at the extreme points $0$ and $\infty $.}  
\begin{equation}
\int\limits_{0}^{\infty }f\left( x\right) g\left( x\right)
xdx=\int\limits_{0}^{\infty }\widetilde{f_{\nu }}\left( s\right) 
\widetilde{g_{\nu }}\left( s\right) sds.  \label{A.3}
\end{equation}
To use (\ref{A.3}) for the calculation of the integral in eq.(\ref{4.5})
we need to know Hankel images of the functions $I_{\nu }(x) K_{\nu}(x)$ 
and $I_{\nu +1}(x) K_{\nu}(x)$ which can be easily obtained from the
general formula \cite{BatErd}
\begin{eqnarray}
K_{-\nu }\left( x\right) I_{\mu }\left( x\right)  &=&
\int\limits_{0}^{\infty}
J_{-\nu +\mu }\left( 2x\sinh t\right) e^{-\left( \nu
+\mu \right) t}dt,  \nonumber \\
- Re\left( \nu +\mu \right)  &<&\frac{3}{2},\quad  Re\left(
-\nu +\mu \right) >-1,  \label{A.4}
\end{eqnarray}
namely
\begin{eqnarray}
\label{A.5}
&&I_{l}\left( x\right) K_{l}\left( x\right) =
\int\limits_{0}^{\infty}
J_{0}\left( xs\right) 
\frac{1}{s\sqrt{s^{2}+4}}e^{-2lt\left( s\right) }sds,  \\
&&I_{l+1}\left( x\right) K_{l}\left( x\right) =
\int\limits_{0}^{\infty}
J_{1}\left( xs\right) 
\frac{1}{s\sqrt{s^{2}+4}}e^{-(2l+1)t\left( s\right) }sds,  
\end{eqnarray}
where $t(s)$ is defined by 
\begin{equation}
2\sinh t=s,\quad dt =\frac{ds}{\sqrt{s^{2}+4}}. 
\label{A.6}
\end{equation}
Though the direct application of eq. (\ref{A.3}) to each term of 
(\ref{4.5}) at first sight seems to be problematic due to the logarithmic 
divergence of the integral over at large $r$, but nevertheless it leads to 
a correct result, because of mutual cancellation of these divergences by 
various terms.

\end{document}